\documentclass[12pt]{article}
\usepackage{graphicx}
\usepackage{epstopdf}
\usepackage{subfig}
\usepackage{caption}
\usepackage{mathrsfs}
\usepackage{amssymb}
\usepackage{amsmath}
\usepackage{verbatim}
\advance\textheight by \topskip
\begin{document}

%\hoffset -0.15cm 
%\tolerance=100000
%\thispagestyle{empty}
%\setcounter{page}{1}

\newcommand{\HPA}[1]{{\it Helv.\ Phys.\ Acta.\ }{\bf #1}}
\newcommand{\AP}[1]{{\it Ann.\ Phys.\ }{\bf #1}}
\newcommand{\be}{\begin{equation}}
\newcommand{\ee}{\end{equation}}
\newcommand{\br}{\begin{eqnarray}}
\newcommand{\er}{\end{eqnarray}}
\newcommand{\ba}{\begin{array}}
\newcommand{\ea}{\end{array}}
\newcommand{\bi}{\begin{itemize}}
\newcommand{\ei}{\end{itemize}}
\newcommand{\bn}{\begin{enumerate}}
\newcommand{\en}{\end{enumerate}}
\newcommand{\bc}{\begin{center}}
\newcommand{\ec}{\end{center}}
\newcommand{\ul}{\underline}
\newcommand{\ol}{\overline}
\def\l{\left\langle}
\def\r{\right\rangle}
\def\as{\alpha_{s}}
\def\ycut{y_{\mbox{\tiny cut}}}
\def\yij{y_{ij}}
\def\epem{\ifmmode{e^+ e^-} \else{$e^+ e^-$} \fi}
\newcommand{\eeww}{$e^+e^-\rightarrow W^+ W^-$}
\newcommand{\qqQQ}{$q_1\bar q_2 Q_3\bar Q_4$}
\newcommand{\eeqqQQ}{$e^+e^-\rightarrow q_1\bar q_2 Q_3\bar Q_4$}
\newcommand{\eewwqqqq}{$e^+e^-\rightarrow W^+ W^-\ar q\bar q Q\bar Q$}
\newcommand{\eeqqgg}{$e^+e^-\rightarrow q\bar q gg$}
\newcommand{\eeqloop}{$e^+e^-\rightarrow q\bar q gg$ via loop of quarks}
\newcommand{\eeqqqq}{$e^+e^-\rightarrow q\bar q Q\bar Q$}
\newcommand{\eewwjjjj}{$e^+e^-\rightarrow W^+ W^-\rightarrow 4~{\rm{jet}}$}
\newcommand{\eeqqggjjjj}{$e^+e^-\rightarrow q\bar 
q gg\rightarrow 4~{\rm{jet}}$}
\newcommand{\eeqloopjjjj}{$e^+e^-\rightarrow q\bar 
q gg\rightarrow 4~{\rm{jet}}$ via loop of quarks}
\newcommand{\eeqqqqjjjj}{$e^+e^-\rightarrow q\bar q Q\bar Q\rightarrow
4~{\rm{jet}}$}
\newcommand{\eejjjj}{$e^+e^-\rightarrow 4~{\rm{jet}}$}
\newcommand{\jjjj}{$4~{\rm{jet}}$}
\newcommand{\qqbar}{$q\bar q$}
\newcommand{\ww}{$W^+W^-$}
\newcommand{\ar}{\rightarrow}
\newcommand{\sm}{${\cal {SM}}$}
\newcommand{\Dir}{\kern -6.4pt\Big{/}}
\newcommand{\Dirin}{\kern -10.4pt\Big{/}\kern 4.4pt}
\newcommand{\DDir}{\kern -8.0pt\Big{/}}
\newcommand{\DGir}{\kern -6.0pt\Big{/}}
\newcommand{\wwqqqq}{$W^+ W^-\ar q\bar q Q\bar Q$}
\newcommand{\qqgg}{$q\bar q gg$}
\newcommand{\qloop}{$q\bar q gg$ via loop of quarks}
\newcommand{\qqqq}{$q\bar q Q\bar Q$}

\def\st{\sigma_{\mbox{\scriptsize t}}}
\def\Ord{\buildrel{\scriptscriptstyle <}\over{\scriptscriptstyle\sim}}
\def\OOrd{\buildrel{\scriptscriptstyle >}\over{\scriptscriptstyle\sim}}
\def\jhep #1 #2 #3 {{JHEP} {\bf#1} (#2) #3}
\def\plb #1 #2 #3 {{Phys.~Lett.} {\bf B#1} (#2) #3}
\def\npb #1 #2 #3 {{Nucl.~Phys.} {\bf B#1} (#2) #3}
\def\epjc #1 #2 #3 {{Eur.~Phys.~J.} {\bf C#1} (#2) #3}
\def\zpc #1 #2 #3 {{Z.~Phys.} {\bf C#1} (#2) #3}
\def\jpg #1 #2 #3 {{J.~Phys.} {\bf G#1} (#2) #3}
\def\prd #1 #2 #3 {{Phys.~Rev.} {\bf D#1} (#2) #3}
\def\prep #1 #2 #3 {{Phys.~Rep.} {\bf#1} (#2) #3}
\def\prl #1 #2 #3 {{Phys.~Rev.~Lett.} {\bf#1} (#2) #3}
\def\mpl #1 #2 #3 {{Mod.~Phys.~Lett.} {\bf#1} (#2) #3}
\def\rmp #1 #2 #3 {{Rev. Mod. Phys.} {\bf#1} (#2) #3}
\def\cpc #1 #2 #3 {{Comp. Phys. Commun.} {\bf#1} (#2) #3}
\def\sjnp #1 #2 #3 {{Sov. J. Nucl. Phys.} {\bf#1} (#2) #3}
\def\xx #1 #2 #3 {{\bf#1}, (#2) #3}
\def\hepph #1 {{\tt hep-ph/#1}}
\def\preprint{{preprint}}

\def\beq{\begin{equation}}
\def\beeq{\begin{eqnarray}}
\def\eeq{\end{equation}}
\def\eeeq{\end{eqnarray}}
\def\a0{\bar\alpha_0}
\def\thrust{\mbox{T}}
\def\Thrust{\mathrm{\tiny T}}
\def\ae{\alpha_{\mbox{\scriptsize eff}}}
\def\ap{\bar\alpha_p}
\def\as{\alpha_{\mathrm{S}}}
\def\aem{\alpha_{\mathrm{EM}}}
\def\b0{\beta_0}
\def\cN{{\cal N}}
\def\cd{\chi^2/\mbox{d.o.f.}}
\def\Ecm{E_{\mbox{\scriptsize cm}}}
\def\ee{e^+e^-}
\def\enap{\mbox{e}}
\def\eps{\epsilon}
\def\ex{{\mbox{\scriptsize exp}}}
\def\GeV{\mbox{\rm{GeV}}}
\def\half{{\textstyle {1\over2}}}
\def\jet{{\mbox{\scriptsize jet}}}
\def\kij{k^2_{\bot ij}}
\def\kp{k_\perp}
\def\kps{k_\perp^2}
\def\kt{k_\bot}
\def\lms{\Lambda^{(n_{\rm f}=4)}_{\overline{\mathrm{MS}}}}
\def\mI{\mu_{\mathrm{I}}}
\def\mR{\mu_{\mathrm{R}}}
\def\MSbar{\overline{\mathrm{MS}}}
\def\mx{{\mbox{\scriptsize max}}}
\def\NP{{\mathrm{NP}}}
\def\pd{\partial}
\def\pt{{\mbox{\scriptsize pert}}}
\def\pw{{\mbox{\scriptsize pow}}}
\def\so{{\mbox{\scriptsize soft}}}
\def\st{\sigma_{\mbox{\scriptsize tot}}}
\def\ycut{y_{\mathrm{cut}}}
\def\slashchar#1{\setbox0=\hbox{$#1$}           % set a box for #1
     \dimen0=\wd0                                 % and get its size
     \setbox1=\hbox{/} \dimen1=\wd1               % get size of /
     \ifdim\dimen0>\dimen1                        % #1 is bigger
        \rlap{\hbox to \dimen0{\hfil/\hfil}}      % so center / in box
        #1                                        % and print #1
     \else                                        % / is bigger
        \rlap{\hbox to \dimen1{\hfil$#1$\hfil}}   % so center #1
        /                                         % and print /
     \fi}                                         %
\def\etmiss{\slashchar{E}^T}
\def\Meff{M_{\rm eff}}
\def\Ord{\lsim}
\def\OOrd{\gsim}
\def\tq{\tilde q}
\def\tchi{\tilde\chi}
\def\lsp{\tilde\chi_1^0}

\def\gam{\gamma}
\def\ph{\gamma}
\def\be{\begin{equation}}
\def\ee{\end{equation}}
\def\bea{\begin{eqnarray}}
\def\eea{\end{eqnarray}}
\def\lsim{\:\raisebox{-0.5ex}{$\stackrel{\textstyle<}{\sim}$}\:}
\def\gsim{\:\raisebox{-0.5ex}{$\stackrel{\textstyle>}{\sim}$}\:}

\def\ino{\mathaccent"7E} \def\gluino{\ino{g}} \def\mgluino{m_{\gluino}}
\def\sqk{\ino{q}} \def\sup{\ino{u}} \def\sdn{\ino{d}}
\def\chargino{\ino{\omega}} \def\neutralino{\ino{\chi}}
\def\cab{\ensuremath{C_{\alpha\beta}}} \def\proj{\ensuremath{\mathcal P}}
\def\dab{\delta_{\alpha\beta}}
\def\zz{s-M_Z^2+iM_Z\Gamma_Z} \def\zw{s-M_W^2+iM_W\Gamma_W}
\def\prop{\ensuremath{\mathcal G}} \def\ckm{\ensuremath{V_{\rm CKM}^2}}
\def\aem{\alpha_{\rm EM}} \def\stw{s_{2W}} \def\sttw{s_{2W}^2}
\def\nc{N_C} \def\cf{C_F} \def\ca{C_A}
\def\qcd{\textsc{Qcd}} \def\susy{supersymmetric} \def\mssm{\textsc{Mssm}}
\def\slash{/\kern -5pt} \def\stick{\rule[-0.2cm]{0cm}{0.6cm}}
\def\h{\hspace*{-0.3cm}}

\def\ims #1 {\ensuremath{M^2_{[#1]}}}
\def\tw{\tilde \chi^\pm}
\def\tz{\tilde \chi^0}
\def\tf{\tilde f}
\def\tl{\tilde l}
\def\ppb{p\bar{p}}
\def\gl{\tilde{g}}
\def\sq{\tilde{q}}
\def\sqb{{\tilde{q}}^*}
\def\qb{\bar{q}}
\def\sqL{\tilde{q}_{_L}}
\def\sqR{\tilde{q}_{_R}}
\def\ms{m_{\tilde q}}
\def\mg{m_{\tilde g}}
\def\Gs{\Gamma_{\tilde q}}
\def\Gg{\Gamma_{\tilde g}}
\def\md{m_{-}}
\def\eps{\varepsilon}
\def\Ce{C_\eps}
\def\dnq{\frac{d^nq}{(2\pi)^n}}
\def\DR{$\overline{DR}$\,\,}
\def\MS{$\overline{MS}$\,\,}
\def\DRm{\overline{DR}}
\def\MSm{\overline{MS}}
\def\ghat{\hat{g}_s}
\def\shat{\hat{s}}
\def\sihat{\hat{\sigma}}
\def\Li{\text{Li}_2}
\def\bs{\beta_{\sq}}
\def\xs{x_{\sq}}
\def\xsa{x_{1\sq}}
\def\xsb{x_{2\sq}}
\def\bg{\beta_{\gl}}
\def\xg{x_{\gl}}
\def\xga{x_{1\gl}}
\def\xgb{x_{2\gl}}
\def\lsp{\tilde{\chi}_1^0}

\def\gluino{\mathaccent"7E g}
\def\mgluino{m_{\gluino}}
\def\squark{\mathaccent"7E q}
\def\msquark{m_{\mathaccent"7E q}}
\def\M{ \overline{|\mathcal{M}|^2} }
\def\utm{ut-M_a^2M_b^2}
\def\MiLR{M_{i_{L,R}}}
\def\MiRL{M_{i_{R,L}}}
\def\MjLR{M_{j_{L,R}}}
\def\MjRL{M_{j_{R,L}}}
\def\tiLR{t_{i_{L,R}}}
\def\tiRL{t_{i_{R,L}}}
\def\tjLR{t_{j_{L,R}}}
\def\tjRL{t_{j_{R,L}}}
\def\tg{t_{\gluino}}
\def\uiLR{u_{i_{L,R}}}
\def\uiRL{u_{i_{R,L}}}
\def\ujLR{u_{j_{L,R}}}
\def\ujRL{u_{j_{R,L}}}
\def\ug{u_{\gluino}}
\def\utot{u \leftrightarrow t}
\def\ar{\to}
\def\sqk{\mathaccent"7E q}
\def\sup{\mathaccent"7E u}
\def\sdn{\mathaccent"7E d}
\def\chargino{\mathaccent"7E \chi}
\def\neutralino{\mathaccent"7E \chi}
\def\slepton{\mathaccent"7E l}
\def\M{ \overline{|\mathcal{M}|^2} }
\def\cab{\ensuremath{C_{\alpha\beta}}}
\def\ckm{\ensuremath{V_{\rm CKM}^2}}
\def\zz{s-M_Z^2+iM_Z\Gamma_Z}
\def\zw{s-M_W^2+iM_W\Gamma_W}
\def\s22w{s_{2W}^2}

\newcommand{\cpmtwo}    {\mbox{$ {\chi}^{\pm}_{2}                    $}}
\newcommand{\cpmone}    {\mbox{$ {\chi}^{\pm}_{1}                    $}}

\begin{flushright}
{SHEP-10-32}\\
\today
\end{flushright}
\vskip0.1cm\noindent
\begin{center}
{{\Large {\bf Higgs phenomenology in the minimal $B-L$
      \\[0.25cm] extension of the Standard Model at LHC}}
\\[1.0cm]
{\large L. Basso, A. Belyaev, S. Moretti and G. M. Pruna\footnote{Speaker}}\\[0.30 cm]
{\it  School of Physics and Astronomy, University of Southampton,}\\
{\it  Highfield, Southampton SO17 1BJ, UK.}
}
\\[1.25cm]
\end{center}

\begin{abstract}
{\small
\noindent
We present some phenomenology of the Higgs sector of the Minimal
$B-L$ $U(1)$ Extension of the Standard Model at the Large Hadron
Collider. In this model, the existence of an extra gauge boson ($Z'$)
and an extra scalar (heavy Higgs) are predicted
as naturally related with the breaking of the $B-L$ (baryon minus
lepton number) symmetry. For this, we have started by deriving the
unitarity bounds in the high energy limit for the Minimal $B-L$ Model
parameter space. This was accomplished by analysing the full class of
Higgs and would-be Goldstone boson two-to-two scatterings at tree
level (exploiting the Equivalence Theorem). Hence, we studied some
peculiar signature that could be observed at the CERN machine in the
search of both light and heavy Higgs bosons.
}

\end{abstract}

%\newpage

%%%%%%%%%%%%%%%%%%%%%%%%%%%%%%%%%%%%%%%%%%%%%%%%%%%%%%%%%%

\section{Introduction}
\label{Sec:Intro}
Despite there is no experimental evidence of a Higgs boson, the Higgs
mechanism is still considered one of the favourite means for
generating the masses of particles. 

In the Standard Model (SM) framework this mechanism is realised by one
Higgs doublet consisting of four degrees of freedom, three of which,
after spontaneous Electro-Weak Symmetry Breaking (EWSB), turn out to
be absorbed in the longitudinal polarisation component of each of
the three weak gauge bosons, $W^{\pm}$ and $Z$, while the fourth one
gives the physical Higgs state $h$.

It is clear that the SM represents a minimal choice that is completely
arbitrary (as far as we know), and in the past years a big effort has
been devoted to explore the implication of more complicated Higgs
models, both in the context of the SM and in Beyond the Standard Model
(BSM) extended theories.

One of the possible BSM scenarios is the minimal $B-L$ (Barion minus
Lepton number) gauge extension
of the SM, which has been recently explored (see \cite{B-L}) as one of the 
candidates in the description of a rather simple phenomenological
framework.

This model has an augmented $B-L$ gauge symmetry, that results in the
natural presence of a new vector boson $Z'$, a new Higgs field
(related to the $B-L$ symmetry breaking) and, in order to preserve the
theory from anomalies, one right-handed neutrino field per
fermionic family (related to three heavy
neutrinos as particle contents of the model).

In the present work we present a brief analysis of the phenomenology
of the $B-L$ Higgs sector at the Large Hadron Collider
(LHC), with emphasis on one distinctive signature of the
model: the heavy neutrino pair production mediated by a light Higgs
in proton-proton collision.

Firstly, in order to give a consistent picture of the allowed
parameter space of the Higgs sector, we
will briefly present the results of
Reference \cite{Basso:2010jt,Basso:2010jm} where the
Higgs parameter space
of the minimal $B-L$ model was studied in detail by accounting for
both experimental and theoretical constraints.

Thereafter, we will present the production cross-sections and
Branching Ratios (BRs), in order to use these results to introduce
some peculiar Higgs signatures at the LHC that are not allowed by the
SM assumption, and therefore they could be the hallmark of the $B-L$
model.

%%%%%%%%%%%%%%%%%%%%%%%%%%%%%%%%%%%%%%%%%%%%%%%%%%%%%%%%%%

\section{The Model}
\label{Sec:Model}
The model under study is the so-called ``pure'' or ``minimal''
$B-L$ model because of the vanishing mixing between the two $U(1)_{Y}$ 
and $U(1)_{B-L}$ groups.
In this model the classical gauge invariant Lagrangian,
obeying the $SU(3)_C\times SU(2)_L\times U(1)_Y\times U(1)_{B-L}$
gauge symmetry, can be decomposed as:
\begin{eqnarray}
\mathscr{L} = \mathscr{L}_{YM} + \mathscr{L}_s + \mathscr{L}_f
+ \mathscr{L}_Y.
\end{eqnarray}

The non-Abelian field strengths in
$\mathscr{L}_{YM}$ are the same as in the SM whereas the Abelian
ones can be intuitively identified.
In this field basis, the co-variant derivative is:
$D_{\mu}\equiv \partial _{\mu} + ig_S
T^{\alpha}G_{\mu}^{\phantom{o}\alpha} + igT^aW_{\mu}^{\phantom{o}a}
+ig_1YB_{\mu} +i(\widetilde{g}Y + g_1'Y_{B-L})B'_{\mu}$.
The ``pure'' or ``minimal'' $B-L$ model is defined by the condition
$\widetilde{g}(EW) = 0$, that implies no mixing between the $Z'$ and the
SM-$Z$ gauge bosons at the tree-level at the EW scale.

The fermionic Lagrangian is the usual $SM$ one, apart from the
presence of Right-Handed (RH) neutrinos. The charges are the usual SM
and $B-L$ ones
(in particular, $B-L = 1/3$ for quarks and $-1$ for leptons). The
$B-L$ charge assignments of the fields as well as the introduction of
new fermionic RH-neutrinos ($\nu_R$) and scalar Higgs ($\chi$, charged
$+2$ under $B-L$) fields are designed to eliminate the triangular
$B-L$ gauge anomalies and to ensure the gauge invariance of the
theory, respectively. Therefore, the $B-L$  gauge extension of the SM
group broken at the Electro-Weak (EW) scale does necessarily require
at least one new
scalar field and three new fermionic fields which are charged with
respect to the $B-L$ group.

The scalar Lagrangian is:
\begin{eqnarray}
\mathscr{L}_s = \left( D^{\mu} H\right)
^{\dagger} D_{\mu}H + \left( D^{\mu} \chi\right) ^{\dagger}
D_{\mu}\chi - V(H,\chi ),
\end{eqnarray}
with the scalar potential given by
\begin{eqnarray}
V(H,\chi ) = m^2H^{\dagger}H + \mu ^2\mid\chi\mid ^2 + \lambda _1
(H^{\dagger}H)^2 +\lambda _2 \mid\chi\mid ^4 + \lambda _3
H^{\dagger}H\mid\chi\mid ^2,
\end{eqnarray}
where $H$ and $\chi$ are the complex
scalar Higgs doublet and singlet fields, respectively.

From this potential, with standard algebraic manipulation (see
\cite{Basso:2010jt}), one finds
the explicit expressions for the Higgs bosons masses and mixing angle
in terms of $\lambda$ parameters.

Being $h_1$ and $h_2$ the scalar fields with masses $m_{h_1}$ and
$m_{h_2}$ respectively (we conventionally choose
$m^2_{h_1} < m^2_{h_2}$), we give the explicit expressions for the
scalar mass eigenvalues:
\begin{eqnarray}\label{mh1}
m^2_{h_1} &=& \lambda _1 v^2 + \lambda _2 x^2 - \sqrt{(\lambda _1 v^2
  - \lambda _2 x^2)^2 + (\lambda _3 xv)^2} \, ,\\ \label{mh2}
m^2_{h_2} &=& \lambda _1 v^2 + \lambda _2 x^2 + \sqrt{(\lambda _1 v^2
  - \lambda _2 x^2)^2 + (\lambda _3 xv)^2} \, ,
\end{eqnarray}
and eigenvectors:
\begin{equation}\label{scalari_autostati_massa}
\left( \begin{array}{c} h_1\\h_2\end{array}\right) =
  \left( \begin{array}{cc}
    \cos{\alpha}&-\sin{\alpha}\\ \sin{\alpha}&\cos{\alpha}
	\end{array}\right) \left( \begin{array}{c}
    h\\h'\end{array}\right) \, ,
\end{equation}
where $-\frac{\pi}{2}\leq \alpha \leq \frac{\pi}{2}$
fulfils:\label{scalar_angle}
\begin{eqnarray}\label{sin2a}
\sin{2\alpha} &=& \frac{\lambda _3 xv}{\sqrt{(\lambda _1 v^2 - \lambda
    _2 x^2)^2 + (\lambda _3 xv)^2}} \, .
\end{eqnarray}

Finally, the Yukawa interactions are: $\mathscr{L}_Y =
-y^d_{jk}\overline {q_{jL}} d_{kR}H - y^u_{jk}\overline {q_{jL}}
u_{kR}\widetilde H - y^e_{jk}\overline {l_{jL}} e_{kR}H
-y^{\nu}_{jk}\overline {l_{jL}} \nu _{kR}\widetilde H -
y^M_{jk}\overline {(\nu _R)^c_j} \nu _{kR}\chi +  {\rm h.c.}$, where
$\tilde H=i\sigma^2 H^*$ and  $i,j,k$ take the values $1$ to $3$,
where the last term is the Majorana contribution and the others the
usual Dirac ones.

%%%%%%%%%%%%%%%%%%%%%%%%%%%%%%%%%%%%%%%%%%%%%%%%%%%%%%%%%%

\section{Results}
\label{Sec:Results}
Firstly, we have made an extensive study on the Higgs-sector
parameter-space allowed by theoretical constraints exploiting the
well-known techniques from consideration on vacuum stability and
triviality (renormalisation group equations (RGEs) techniques, see
\cite{Basso:2010jm}) and
perturbative unitarity (PU) (see \cite{Basso:2010jt}). The latter, in
particular, is not energy-scale
dependent, and it results in a simpler description of the
$m_{h_1}-m_{h_2}-\alpha$ allowed space, and we will mainly consider it
in the following analysis.

By analysing the full class of Higgs and would-be Goldstone boson
two-to-two scatterings at tree level (exploiting the Equivalence
Theorem) one finds the following result: the theory is PU-stable if
$m_{h_1}<700$ GeV and
\begin{eqnarray}
m_{h_2}<2\sqrt{\frac{2}{3}}\ {\rm min} \left( \frac{m_W}{\sqrt{\alpha_W}
    \sin{\alpha}},\sqrt{\pi}x\right),
\end{eqnarray}
where $\alpha_W=\alpha_{em}/\sin^2\theta_W$.

Considering the experimental limits from LEP (which established that
the safest choice for the light Higgs boson mass is $m_{h_1}>115$ GeV)
we have a complete definition of the allowed Higgs-sector parameter
space.
                    
For completeness, we have also combined the RGEs and PU techniques in
order to find a dynamical constraint on the $g'_1$ domain (see
\cite{Basso:2010hk}). From this method, assuming that the $B-L$
symmetry breaking occurs at the TeV scale, one finds that the model is
PU-stable to the Planck scale only if $g'_1<0.23$.

Thereafter, we have analysed the production mechanism channels both for
$h_1$ and $h_2$ in proton-proton colliders, with emphasis on two
LHC energy-luminosity configurations: ``early discovery scenario''
with $\sqrt{s}=7$ GeV and $L=1$ ${\rm fb}^{-1}$, ``full integrated
luminosity'' scenario with $\sqrt{s}=14$ GeV and $L=300$ ${\rm
  fb}^{-1}$.

\begin{figure}[h]
\begin{minipage}{14pc}
\includegraphics[width=14pc]{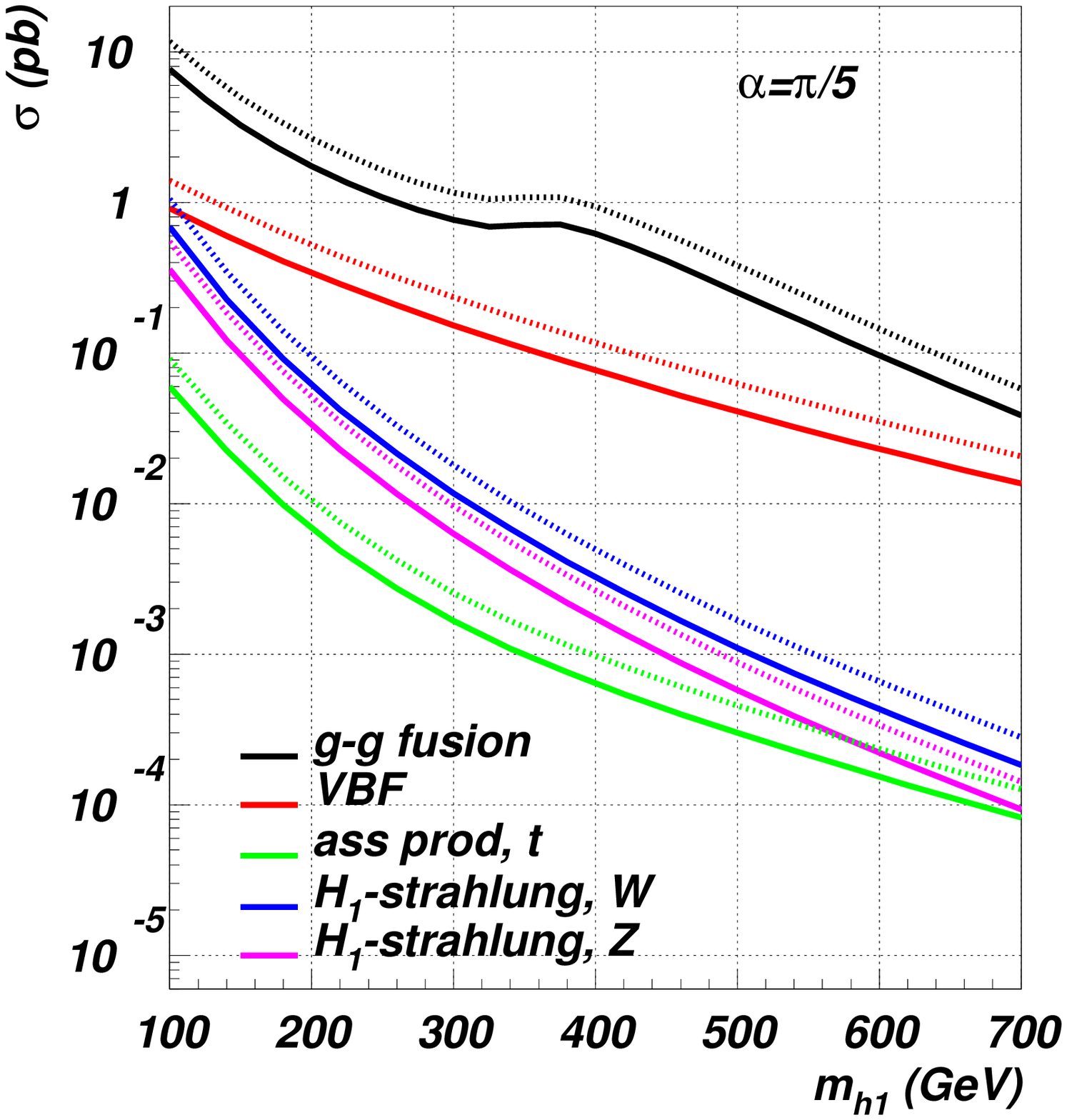}
\caption{\label{merged}Cross-section for $h_1$ production at LHC
  ($\sqrt{s}=7$ TeV) plotted against
  the light Higgs boson mass $m_{h_1}$ at $\alpha=\pi/5$ in $g$-$g$
  fusion (black line), VBF (red
  line), Higgs-strahlung from $W^{\pm}$ (blue line) and
  $Z$ (violet line) and  associated production with top (green
  line). The 
  SM-like case has been superimposed in dotted lines.}
\end{minipage}\hspace{2pc}%
\begin{minipage}{14pc}
\includegraphics[width=14pc]{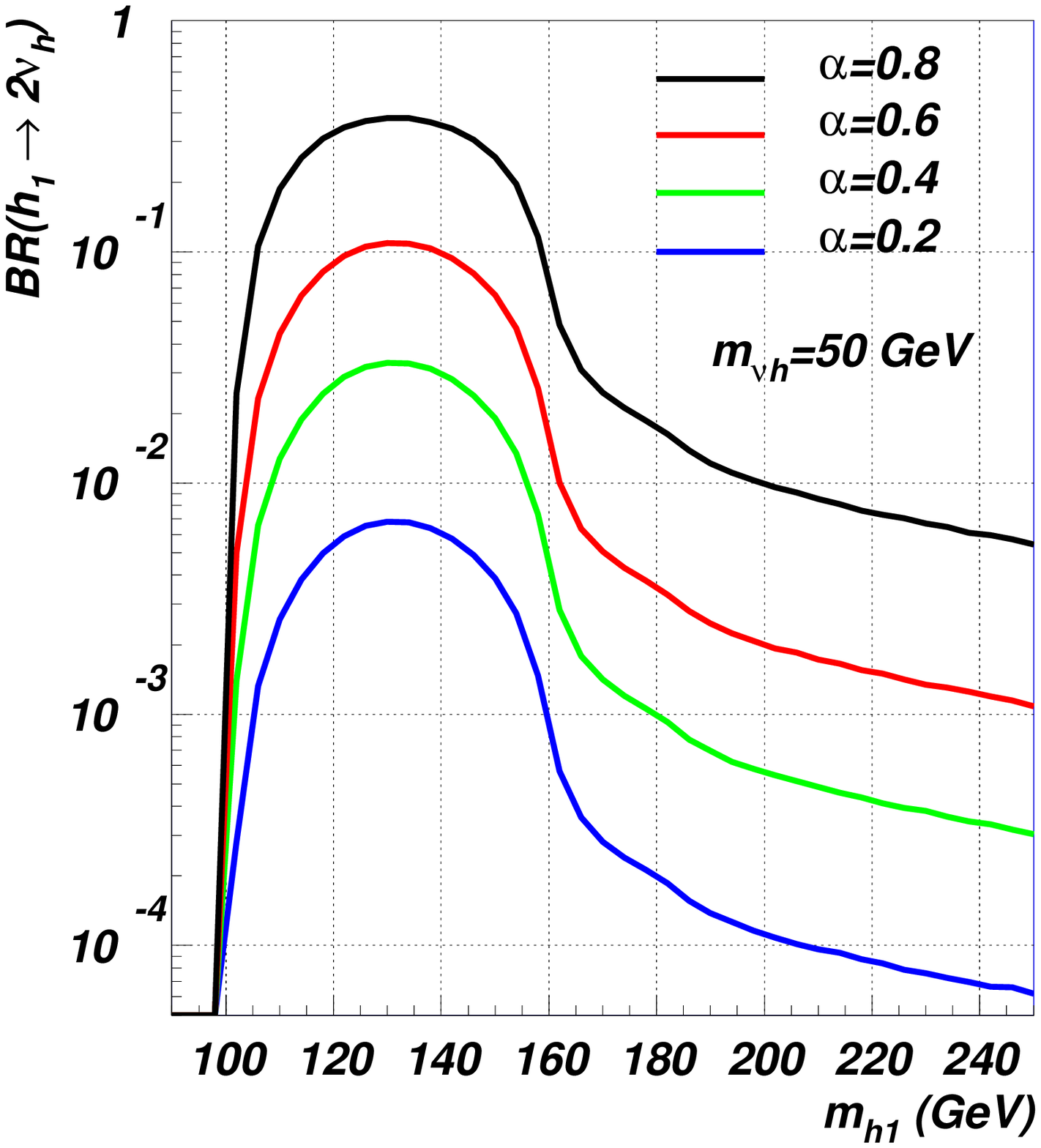}
\caption{\label{branching}Branching ratio for the light Higgs boson
  decaying in two heavy neutrinos ($h_1\rightarrow 2\nu_h$) in the
  minimal $B-L$ model plotted against the
  light Higgs boson mass $m_{h_1}$ for $\nu_h=50$ GeV and several
  values of the mixing angle $\alpha$ in units of $\pi/2$:
  $\alpha=0.2$ (blue line), $\alpha=0.4$ (green line), $\alpha=0.6$
  (red line), $\alpha=0.8$ (black
  line).}
\end{minipage} 
\end{figure}

As explicit example, in figure \ref{merged} we show the full set of
production mechanisms for $h_1$ in proton-proton collision at
$\sqrt{s}=7$ GeV for $\alpha=\pi/5$: as in the SM case, the dominant
mode is represented by the gluon-gluon fusion (black line) process,
while the inclusive processes as vector boson fusion (VBF) (red line),
the H-strahlung processes 
(blue line for $W^{\pm}$ and violet line for $Z$) and the associated
production of top and Higgs (green line) represent a significantly
smaller contribution. For completeness, we have superimposed the
SM-like case ($\alpha=0$) in dotted lines.

Besides, we have evaluated the BRs both for
$h_1$ and $h_2$, in the search for configurations in which it could be
possible to have any peculiar signature of the model.

Considering the $h_1$-decay, we have analysed the role
that a ``light'' heavy neutrino mass ($m_Z/2<m_{\nu_h}<m_W$) could
play in the BRs, in order to establish if such decaying-channel could
be visible in the early discovery scenario at LHC. In
particular, in figure \ref{branching} we plot the Branching
ratio for $h_1\rightarrow 2\nu_h$ (summing over the three generation of
heavy neutrinos) against the light Higgs boson mass
$m_{h_1}$ for $\nu_h=50$ GeV\footnote{We assume that the three heavy
  neutrinos mass eigenstates are degenerate as this does not affect
  our analysis.} and several values of the mixing angle
$\alpha$ in units of $\pi/2$: $\alpha=0.2$ (blue line), $\alpha=0.4$
(green line), $\alpha=0.6$ (red line), $\alpha=0.8$ (black line).

Considering the $h_2$-decay, instead, we have analysed how the
presence of a ``light'' $Z'$-boson (the $m_{Z'}=210$ GeV choice forces
us to choose $g'_1<0.03$ because of the LEP limits, see
\cite{Cacciapaglia:2006pk}) could
affect the BRs, in order to establish if such decaying-channel could
be visible in the full integrated luminosity scenario at
LHC. Nevertheless, we have also studied the role of a light Higgs
boson ($m_{h_1}=120$ GeV) in the $h_2$'s BRs, because of the fact that
the channel $h_2\rightarrow 2h_1$ represents a peculiar signature of
the model, that is distinctive with respect to supersymmetrical Models
in which the mixing in the scalar sector is forbidden.

Finally, by combining the two analysis (Higgses productions and BRs),
we have made a detailed study of the cross-section for some peculiar
signature of the model: $pp\rightarrow h_1 \rightarrow \nu_h\nu_h$,
$pp\rightarrow h_2 \rightarrow h_1h_1$ and $pp\rightarrow h_2
\rightarrow Z'Z'$. The analysis of each of these processes has shown
how there is the possibility to observe such signatures both in the
early discovery scenario ($pp\rightarrow \nu_h\nu_h$) and
in the full integrated luminosity scenario ($pp\rightarrow h_1h_1$ and
$pp\rightarrow Z'Z'$).

\begin{figure}[h]
\includegraphics[width=14pc]{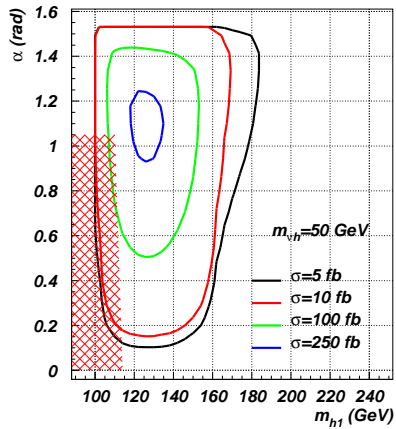}\hspace{2pc}%
\begin{minipage}[b]{14pc}\caption{\label{contour}Cross-section contour 
    plot for the process $pp\rightarrow 
    h_1\rightarrow \nu_h \nu_h$ at LHC $\sqrt{s}=7$ TeV, plotted
    against $m_{h_1}$-$\alpha$, with $m_{\nu_h}=50$
    GeV. Several values of the
    cross-section have been considered: $\sigma=5$ fb (black line),
    $\sigma=10$ fb (red line), $\sigma=100$ fb (green line),
    $\sigma=250$ fb (blue line). The red-shadowed region is excluded
    by the LEP experiments.}
\end{minipage}
\end{figure}

In particular, in figure \ref{contour} we show the explicit result for
the $pp\rightarrow h_1\rightarrow \nu_h \nu_h$ process at LHC with
$\sqrt{s}=7$ TeV and $m_{\nu_h}=50$ GeV: a cross-section contour
``sliced'' in the $m_{h_1}$-$\alpha$ plane. Several values of the
cross-section have been considered: $\sigma=5$ fb (black line),
$\sigma=10$ fb (red line), $\sigma=100$ fb (green line), $\sigma=250$
fb (blue line). The red-shadowed region is excluded by the LEP
experiments.

Even if we consider a low-luminosity scenario
($L\simeq 1$ fb$^{-1}$), it is clear from the plot that there is a
noticeable allowed parameter space for which the rate of events is
considerably large: when the integrated luminosity reaches $L=1$
fb$^{-1}$ (that is equivalent to $18-24$ months of $\sqrt{s}=7$ TeV
running at LHC according to the official programme) we estimated a
collection of $\sim 10$ heavy neutrino pair 
productions for $100$ GeV$<m_{h_1}<165$ GeV and
$0.05\pi<\alpha<0.48\pi$, that scales up to $\sim 10^2$ events for
$110$ GeV$<m_{h_1}<150$ GeV and $0.15\pi<\alpha<0.46\pi$.

This represents a clear chance to establish a heavy neutrino discovery
within the next two years at the CERN machine.

%%%%%%%%%%%%%%%%%%%%%%%%%%%%%%%%%%%%%%%%%%%%%%%%%%%%%%%%%%

%\section*{References}

\end{document}